# Elemental nitrogen partitioning in dense interstellar clouds.

Julien Daranlot,[a,b] Ugo Hincelin,[c,d] Astrid Bergeat,[a,b] Michel Costes,[a,b] Jean-Christophe Loison,[a,b] Valentine Wakelam,[c,d] Kevin M. Hickson[a,b,1]

[a]Univ. Bordeaux, ISM, UMR 5255, F-33400 Talence, France.
[b]CNRS, ISM, UMR 5255, F-33400 Talence, France.
[c]Univ. Bordeaux, LAB, UMR 5804, F-33270 Floirac, France.
[d]CNRS, LAB, UMR 5804, F-33270 Floirac, France.

**Many chemical models of dense interstellar clouds predict that the majority of gas-phase elemental nitrogen should be present as $N_2$, with an abundance approximately five orders of magnitude less than that of hydrogen. As a homonuclear diatomic molecule, $N_2$ is difficult to detect spectroscopically through infrared or millimetre-wavelength transitions so its abundance is often inferred indirectly through its reaction product $N_2H^+$. Two main formation mechanisms each involving two radical-radical reactions are the source of $N_2$ in such environments. Here we report measurements of the low temperature rate constants for one of these processes, the N + CN reaction down to 56 K. The effect of the measured rate constants for this reaction and those recently determined for two other reactions implicated in $N_2$ formation are tested using a gas-grain model employing a critically evaluated chemical network. We show that the amount of interstellar nitrogen present as $N_2$ depends on the competition between its gas-phase formation and the depletion of atomic nitrogen onto grains. As the reactions controlling $N_2$ formation are inefficient, we argue that $N_2$ does not represent the main reservoir species for interstellar nitrogen. Instead, elevated abundances of more labile forms of nitrogen such as $NH_3$ should be present on interstellar ices, promoting the eventual formation of nitrogen-bearing organic molecules.**

\body

## Introduction

Elemental nitrogen is present in the atmospheres of telluric planets predominantly in the form of $N_2$. As a stable molecule with a high bond energy, it is hard to break N out of $N_2$, so the amount of nitrogen available for the formation of complex prebiotic molecules in such environments depends on the presence of nitrogen in more labile forms such as $NH_3$. Despite its high abundance in the Solar System, observations of $N_2$ beyond our own planetary system are scarce, as it possesses no allowed rotational or vibrational transitions. Only one direct measurement (1) of $N_2$ in the far ultraviolet wavelength range (thereby accessing its electronic transitions) has been reported in a diffuse interstellar cloud, where low column densities allow light to pass and a star along the line of sight was used to probe the species within. In dense molecular clouds where temperatures as low as 10 K prevail, UV starlight is absorbed long before it can penetrate to the cloud core where $N_2$ is predicted to form efficiently (2, 3). Instead, gas-phase $N_2$ densities are inferred through observations of $N_2H^+$, product of the $N_2 + H_3^+$ reaction (4-6). The major pathways for $N_2$ formation in dense clouds rely on reactions involving neutral radical species:

| | | |
|---|---|---|
| N + OH → NO + H | 1 | |
| N + NO → N$_2$ + O | 2 | Mechanism (I) |
| | | |
| N + CH → CN + H | 3 | |
| N + CN → C + N$_2$ | 4 | Mechanism (II) |

Recent experimental and theoretical studies of reactions 1 (7) and 2 (8-10) indicate that these reactions are slower than previously thought, suggesting that current astrochemical models overestimate molecular nitrogen abundances produced by mechanism (I) both at steady state and at specific times. The influence of mechanism (II) on interstellar N$_2$ abundances is currently unknown given that neither reaction 3 nor reaction 4 has been studied at temperatures lower than 200 K. Nevertheless, both of these reactions are estimated to react rapidly, even at 10 K.

Unfortunately, the measurement of rate constants for reactions at temperatures below 100 K is far from trivial. Cryogenic cooling methods can be used to attain such temperatures, however these techniques are hampered by the condensation of reactants and precursor molecules onto the reactor walls. As a result, the range of reactions that can be studied is restricted to those between species with substantial vapor pressures at low temperature. During the early 1990s, the CRESU technique (Cinetique de Reaction en Ecoulement Supersonique Uniforme or reaction kinetics in a uniform supersonic flow) that had already been applied to the study of ion-molecule reactions (11) was adapted to study reactions between neutral species at low temperatures (12). This development was largely responsible for proving that reactions between neutral species play an important role in cold regions of interstellar space. The CRESU method uses Laval nozzle expansions to produce cold supersonic flows with a uniform density and temperature profile for a period of several hundreds of microseconds allowing fast reactions to be studied. As the cold jet of gas is isolated from the reactor walls, problems with reagent condensation are avoided. This technique has so far led to the measurement of rate constants for more than 100 radical-molecule reactions to temperatures as low as 6 K (13). In stark contrast, few low temperature investigations of reactions involving two unstable species have been reported (7, 14). These reactions are particularly challenging for experimentalists, requiring the production of one of the radicals in excess (for the pseudo-first-order kinetic analysis to be valid) and the measurement of its concentration for absolute rate constants to be extracted. As a result, current models rely heavily on calculations, extrapolations or estimates to predict their rates. The recent development of the discharge-flow CRESU technique (8) and its application to the study of unstable radical reactions through the relative rate method (7) has furnished low temperature rate constants for reactions 1 and 2 whilst demonstrating that studies of other atom – radical reactions are entirely feasible. Here we report measurements of the rate constant for reaction 4 from 296 K to 56 K using this method. The influence of the new rate constants for reactions 1, 2 and 4 on elemental nitrogen partitioning in dense interstellar clouds and in prestellar cores is tested for the first time using chemical models incorporating both gas-phase and grain surface chemistry. Details of both the experimental method and the model are provided in the *Materials and Methods* section whereas the experimental and model cross-checks performed for this study are outlined in the SI Text.

**Results and Discussion**

The measured temperature dependent rate constants for the N + CN reaction are displayed in Fig. 1 alongside earlier experimental work (15, 16) and the recommended values as proposed by the Kinetic Database for Astrochemistry (KIDA) (17). The measured rate constants are also listed in Table S1 alongside other relevant information. The first step in the measurement of such quantities relies on the application of the pseudo-first-order approximation, where one of the reagent species is held in a large excess so that its concentration remains effectively constant throughout the reaction, thereby facilitating the kinetic analysis. Under circumstances where the excess reagent is a stable molecule, it is a trivial procedure to determine its concentration from its partial pressure. A linear least squares fit to the resulting pseudo-first-order rate constants plotted as a function of the coreagent concentration yields the second-order rate constant directly. For the present experiments however, where the atomic nitrogen concentration, [N], can no longer be determined by this method, rate constants for the N + CN reaction were measured by the relative rate method using the N + OH reaction as the reference process, both reactions occurring concurrently in the same reactor. Fig. 2 shows the temporal evolution of reactant CN and reference OH fluorescence signals recorded simultaneously in the presence of excess atomic nitrogen at 296K and 56 K.

Under these conditions, both reactants decay exponentially to zero, with a characteristic decay time, $\tau$, proportional to the product of [N], and the rate constant $k_1$ or $k_4$. As any single pair of OH and CN decays is measured with the same [N], the difference in $\tau$ directly reflects the differing reactivity of the two reactions. Given that [N] was held in large excess with respect to the OH and CN radical concentrations for all experiments, simple exponential fits to the OH and CN fluorescence signals yield the pseudo-first-order rate constants $k'_1 \equiv k_1[N]$ and $k'_4 \equiv k_4[N]$. This procedure was performed at several values of [N] by varying the $N_2$ flow and/or the power of the microwave discharge. The values of $k'_4$ obtained in this way for a specified temperature were plotted as a function of the corresponding values of $k'_1$. A weighted linear least squares fit to the data yields the ratio of the two rate constants $k_4 / k_1$ from the slope, an example of which is shown in Fig. 3.

The temperature dependent rate constants for reaction 4 presented in Fig.1 were finally obtained by multiplying the ratio obtained from the slope of plots similar to the one presented in Fig. 3 by the value of the rate constant previously obtained for reaction 1 at the corresponding temperature. The rate of reaction 1 is approximately constant over the range 296 K – 56 K (7) with $k_1 \approx 4.5 \times 10^{-11}$ cm$^3$ s$^{-1}$ so comparison between the pairs of decays in panels (A) and (B) of Fig. 2 clearly demonstrates that the reactivity of reaction 4 falls as the temperature is lowered. At 296 K, $k_4 = 9 \times 10^{-11}$ cm$^3$ s$^{-1}$, in good agreement with earlier measurements (15, 16) and this value falls to $5 \times 10^{-11}$ cm$^3$ s$^{-1}$ at 56 K.

A simple A $\times$ $(T/300)^B$ fit to the data yields a temperature dependence $T^{(0.42 \pm 0.09)}$ (A = (8.8 ± 3.8) $\times 10^{-11}$ cm$^3$ s$^{-1}$). As the rate constant varies only weakly with temperature, extrapolation allows us to predict a value of $k_4(10\ K) = 2 \times 10^{-11}$ cm$^3$ s$^{-1}$, more than an order of magnitude smaller than the value of $3 \times 10^{-10}$ cm$^3$ s$^{-1}$ recommended by current astrochemical databases UMIST (19) and OSU 2008 (Herbst, E. Ohio State University Astrochemical Database (update OSU_09_2008) http://www.physics.ohio-state.edu/~eric/research_files/osu_09_2008).

The recently measured and calculated rate constants for reactions 1 (7), 2 (8-10) and 4 might influence the $N_2$ abundance predicted by chemical models of dark clouds. To test their importance to $N_2$ formation, we used the chemical model Nautilus (20) which solves the kinetic equations for both gas-phase and grain surface chemistry. This model incorporates gas-phase reactions, the sticking of gas-phase species onto interstellar grains, the evaporation of surface species into the gas-phase and chemical reactions occurring at the surface of the grains. Typical dense cloud conditions were used and species were initially present in their atomic or ionic form at their diffuse cloud abundances as given in Table S2, assuming that dense molecular clouds form from diffuse media, except for hydrogen, which was present as $H_2$. The model runs were initially performed with a depleted oxygen abundance compared to that of carbon as suggested by Jenkins (21) and Whittet (22), yielding an elemental C/O ratio of 1.2 (see the SI Text for further details). Depleted elemental oxygen was required to reproduce the observed low gas-phase abundance of $O_2$ in dense clouds in a recent modeling study (20). However, given the probable model dependence on initial elementary abundances which are themselves poorly constrained, secondary model runs were also undertaken in which the elemental C/O ratio was fixed to 0.7 using a larger oxygen abundance of $2.4 \times 10^{-4}$ with respect to total hydrogen ($n_H = n(H) + 2n(H_2)$) to test the effect of excess elemental oxygen. The gas and dust temperature was fixed at 10 K and a visual extinction ($A_V$) of 10, a cosmic-ray ionization rate of $1.3 \times 10^{-17}$ s$^{-1}$ and a total hydrogen density of $2 \times 10^4$ cm$^{-3}$ were used. The model follows species in the gas-phase and on interstellar grains as a function of time. The gas-phase chemical network used was updated in 2011 according to recommendations made by experts contributing to KIDA (17), and should be the most representative set of input parameters currently available for modeling purposes.

The chemical model was run using three sets of kinetic input parameters: a) the most recently recommended rate constants by the OSU database (OSU_09_2008) ; b) the KIDA recommended values ; c) the KIDA recommended values replacing rate constants for reactions 1, 2 and 4 with the new values of $k_1 = (2.5 \pm 1.0) \times 10^{-11}$, $k_2 = (1.0^{+1.0}_{-0.5}) \times 10^{-11}$ and $k_4 = (2.0 \pm 1.0) \times 10^{-11}$ cm$^3$ s$^{-1}$ from extrapolation of the experimental data to 10 K. The quoted errors on these rate constants were estimated from the spread of the available theoretical and experimental data. The percentage of elemental nitrogen that forms $N_2$ in these models for a C/O ratio of 1.2 are shown in Fig. 4 (**A**). The corresponding models for a C/O ratio of 0.7 are shown in Fig 4 (**C**).

$N_2$ formation is already less efficient for both high and low C/O ratios if we compare the results of models a) and b) as a result of the numerous updates of the original OSU scheme proposed by the KIDA experts. The inclusion of new rate constants for reactions 1, 2 and 4 in model c) reduces $N_2$ yields even further. For model c), 40 % of the initial elemental nitrogen at most is converted to $N_2$ (in the gas-phase and on grain surfaces) for high C/O ratios and an even smaller value of 17 % is obtained for the low C/O ratio case.

Figs. 4 (**B**) and (**D**) show $N_2H^+$ fractional abundances ($/n_H$) (for high and low C/O ratios respectively with model c), alongside the observational constraints on $N_2H^+$ from Taurus Molecular Cloud (TMC-1, CP Peak) of $1.5 \times 10^{-10}$ (5) and $3.5 \times 10^{-10}$ (23). The model c) abundances of gas-phase N, $N_2$ and $NH_3$ ice are also shown.

At times of $2 \times 10^5$ and $(3-5) \times 10^5$ yr, modeled $N_2H^+$ abundances are in good agreement with the observed ones, providing the age constraints for our high and low C/O ratio models respectively. At these times, $N_2$ has a predicted gas-phase abundance of $(6 \pm 2) \times 10^{-6}$ ($/n_H$) for the low oxygen case and $(4 \pm 2) \times 10^{-6}$ ($/n_H$) for the high oxygen case amounting to only $(21 \pm 6)$ % and $(12 \pm 5)$ % respectively of the total initial nitrogen. Errors on these gas phase $N_2$ values were obtained by performing test model runs with rate constants for reactions 1, 2 and 4 at the limits given by our estimated uncertainties. Instead of producing $N_2$, atomic nitrogen depletes onto interstellar grains, where $NH_3$ ice forms efficiently in both low oxygen and high oxygen models. At long times, the condensed phase abundance of $N_2$ falls (> $10^6$ yr) as a result of dissociation by cosmic rays, slowly converting it to $NH_3$.

Previous studies have suggested that $N_2$ is mostly produced by mechanism (I) in dark clouds (5, 23). Our analysis suggests that this scheme might be only a minor source of gas-phase $N_2$ compared with mechanism (II). This can be highlighted by the results of the low C/O ratio model where we would expect the presence of excess atomic oxygen to increase the influence of mechanism (I) by promoting OH formation through a series of ion-neutral reactions with $H_3^+$. In this case, the reactive fluxes of reactions 2 and 4 (defined as the product $k$[A][B] where [A] and [B] are the densities of reactants A and B respectively) from (2-5) × $10^5$ yr are similar, producing equivalent quantities of $N_2$ (see Fig. S1(**A**)). In contrast, the reactive flux of reaction 4 is 20 times greater than for reaction 2 in the high C/O ratio model over the same time period resulting in a higher overall $N_2$ yield. Clearly, the updated scheme (I) is now an inefficient $N_2$ production mechanism. To rationalize the model results, we can examine the abundances of the relevant precursor species CN and NO (see Fig S1 (**B**)). CN abundances are reduced by a factor of 10 for the low C/O case as the O + CN reaction (with a recommended rate constant (17) of $4 \times 10^{-11}$ $cm^3$ $s^{-1}$ at 10 K) becomes the major CN removal mechanism. In the absence of an efficient removal mechanism for NO (the equivalent O + NO reaction does not occur), its abundance rises by a factor of 10 without significantly increasing $N_2$ yields. One factor that has so far been overlooked is the importance of reaction 3. If the rate constant at 10 K is less than the currently predicted value of $2.3 \times 10^{-10}$ $cm^3$ $s^{-1}$, the $N_2$ yield of mechanism (II) would decrease by a corresponding amount, due to the reduced reactive flux of reaction 4. Future low temperature measurements of the rate for reaction 3 are required to resolve this issue and to provide a definitive answer as to which of the two mechanisms is really dominant. In either case, the clear conclusion is that total $N_2$ production falls.

It is interesting to test the effect of our newly determined rate constants on $N_2$ formation in other more dense regions of interstellar space such as prestellar cores; the evolutionary stage following dense clouds but preceding protostar formation. The nitrogen chemistry of prestellar cores, has been the subject of debate in recent years, because models of such regions employing recommended rate constants have been unable to reproduce observed abundances of nitrogen containing species (24-26). One of the most well-known prestellar cores is Barnard 68 (B68). The density structure of B68 observationally determined by Alves et al. (27) was used to compute the abundances of the major nitrogen bearing species as a function of visual extinction Av, which can be related to the radius towards the core center. More details on the prestellar core model parameters can be found in the Materials and Methods section. The predicted abundances of the major nitrogen bearing species are presented in Fig. 5 (**A**) for a time of $5 \times 10^6$ yr whilst the results of the time dependent model are shown in Fig. 5 (**B**). The predicted abundances of nitrogen bearing species as a function of Av are very similar to the earlier results of Maret et al. (6). However, the initial conditions used in our model (species in atomic form) did not need to be modified to obtain these results. Moreover, tests in which some of the initial elements were put in molecular form at the outset (notably with 50 % of the available carbon in the form of CO) yielded essentially the same fractional abundances (see the SI Text for more details). Mechanistically, the lower rate constants for our updated scheme (I) lead to less gas phase $N_2$ production and a correspondingly low $N_2H^+$ abundance. In contrast, Maret et al. (6) adjusted the initial fraction of elemental nitrogen present as $N_2$ (thereby allowing $N_2H^+$ formation) whilst blocking the time dependent formation of $N_2$ through low gas phase atomic oxygen abundances, effectively offsetting the previously high rate constants for reactions 1 and 2. In a similar manner to our dense cloud results presented in Fig. 4, an increase in initial elemental oxygen (with C/O = 0.7) under prestellar core conditions actually results in even lower total $N_2$ yields (see Fig. S2).

In the absence of efficient gas-phase removal mechanisms for atomic nitrogen in our updated model, these atoms deplete onto interstellar grains where surface reactions with atomic hydrogen are predicted to transform them to $NH_3$ ices. This result is supported by observations of dense regions which have so far failed to detect $N_2$ ices (28), whilst observations of gas-phase $NH_3$ originating from the sublimation of icy mantles (29) or direct observations of $NH_3$ ices in young stellar objects (YSOs) (30) indicate that as much as 20 % of total elemental nitrogen could be present as condensed phase $NH_3$. Our model predicts that a larger value of 45% of the elemental nitrogen should be present as $NH_3$ ice (from model c) results with a C/O ratio of 0.7) in TMC-1 at times when the $N_2H^+$ abundance is well reproduced. The remaining elemental nitrogen at these times is partitioned between gas-phase N, gas-phase $N_2$ and $N_2$ ices which are themselves converted to $NH_3$ ice on longer timescales. The predicted amount of $NH_3$ ice is also substantially larger than the observed one in quiescent dark clouds (5% compared to $H_2O$ ices) (31). The only surface reaction mechanism currently considered in our model is the Langmuir-Hinshelwood one whereby two thermalized species are allowed to diffuse across the surface until they react. The inclusion of other surface processes could allow the formation of more complex N-bearing species or even maintain nitrogen in its atomic form. Indeed, the observation of non-hydrogenated surface species, such as $CO_2$ and OCS (32) does indicate that mechanisms other than the diffusive one do exist (33). The predicted $NH_3$ ice abundances are also much larger than the ones observed in comets (34). As in interstellar ices, $NH_3$ is the most abundant N-bearing species observed in comets but it only accounts for a small fraction of the total elemental nitrogen, whose reservoir is currently unknown. Although our study cannot give a definitive answer to this question, the use of lower rate constants for reactions (1), (2) and (4) whilst considering depletion onto grain surfaces, indicates that the ultimate reservoir of nitrogen in interstellar environments is unlikely to be $N_2$ either in the gas-phase or on the grain surfaces. This result has potential consequences for our view of the development of the early Earth, which probably possessed a neutral atmosphere consisting of mostly $CO_2$ and $N_2$. As considerable energy is required to break N out of $N_2$, other nitrogen bearing molecules such as ammonia in interplanetary dust from the primitive nebula have been proposed as possible sources of fixed nitrogen easily usable by the first forms of life on Earth (35).

**Materials and Methods**

**Experimental.** Experiments were performed using a miniaturized continuous supersonic flow reactor modified from the original apparatus designed by Rowe et al. (11) The main features of the present system have been previously described (7). Briefly, four Laval nozzles were used to perform kinetic measurements at specified temperatures of 56 K, 87 K, 152 K and 169 K. Measurements at 296 K were performed in the reactor by removing the nozzle and by significantly reducing the flow velocity to eliminate pressure gradients in the observation region. Ground-state atomic nitrogen was generated upstream of the Laval nozzle by the microwave discharge technique. A Vidal type microwave discharge cavity (36) operating at 2.45 GHz and up to 200 W was mounted on one of the quartz inlet tubes entering the Laval nozzle reservoir. In this way, excess concentrations of ground state atomic nitrogen estimated to be as high as $3.3 \times 10^{14}$ cm$^{-3}$ could be produced. An earlier study (7) showed that atomic nitrogen in excited states $N(^2P^0)$ and N($^2$D$^0$) was negligible in the supersonic flow under similar conditions. The high discharge power used to produce excess atomic nitrogen also resulted in elevated gas temperatures within the nozzle reservoir. Moreover, the reservoir temperature was also found to vary as a function of the molecular nitrogen flow through the discharge during some experiments. As a result, for a specified nozzle, the discharge power was varied for different molecular nitrogen flows to maintain a constant reservoir temperature, as verified by a type K thermocouple inserted into the reservoir prior to the experiments. At the same time, the supersonic flow was characterized through the measurement of the impact pressure using a Pitot tube to verify that the microwave discharge did not perturb the downstream flow. The measured reservoir temperatures, impact and stagnation pressures were subsequently employed in the flow calculations.

OH radicals in the X$^2\Pi_i$ state were generated by the in-situ pulsed photodissociation of H$_2$O$_2$ using the unfocused output of a 10 Hz frequency quadrupled Nd:YAG laser at 266 nm with ~ 22 mJ of pulse energy. H$_2$O$_2$ was introduced into the flow by bubbling a small flow of the carrier gas through a 50% weight mixture of H$_2$O$_2$ / H$_2$O. An upper limit of $3.6 \times 10^{13}$ cm$^{-3}$ was estimated for the gas phase concentration of H$_2$O$_2$ in the supersonic flows and $5.0 \times 10^{13}$ cm$^{-3}$ at 296 K from its saturated vapour pressure, providing OH concentrations lower than $4.3 \times 10^{11}$ cm$^{-3}$ in the supersonic flows and $6.0 \times 10^{11}$ cm$^{-3}$ at 296 K from calculations of the photodissociation efficiency of H$_2$O$_2$. Simultaneously, CN radicals in the X$^2\Sigma^+$ state were generated by the 266 nm photolysis of ICN molecules entrained in the flow by passing a small flow of carrier gas over a sample of solid ICN held in a separate container. An upper limit of $2.1 \times 10^{12}$ cm$^{-3}$ was estimated for the gas phase concentration of ICN in the supersonic flows and $5.2 \times 10^{12}$ cm$^{-3}$ at 296 K from its saturated vapour pressure. CN concentrations in the supersonic flow and at 296 K were estimated to be $8.5 \times 10^{10}$ and $2.0 \times 10^{11}$ cm$^{-3}$ respectively. From the estimated minor reagent densities, reaction between CN and OH should not occur to any great extent in the flow. Nevertheless, test experiments were performed which showed that OH decays recorded in the absence of CN yielded identical first-order rate constants to those obtained when CN was present in the flow, confirming this hypothesis.

The probe laser system for the detection of OH radicals by laser-induced fluorescence (LIF) has been described previously (37). CN radicals were followed by LIF via the R-branch lines of the (1, 0) B$^2\Sigma^+$ ← X$^2\Sigma^+$ band. For this purpose, the third harmonic 355 nm radiation of a pulsed single longitudinal mode Nd:YAG laser was used to pump an OPO system to produce tuneable radiation around 538 nm. This radiation was sum frequency mixed with the residual fundamental radiation of the same Nd:YAG laser at 1064 nm in a BBO crystal to produce tuneable radiation around 357.55 nm. CN radicals were observed via the B$^2\Sigma^+$ → X$^2\Sigma^+$ (1, 1) band using a second UV sensitive PMT coupled with a 10 nm FWHM interference filter centred on 390 nm and a second boxcar integrator. Both probe lasers were coaligned and counterpropagated with respect to the photolysis laser in the reactor.

LIF signals from the reactant OH and CN radicals were recorded simultaneously. For a given atomic nitrogen concentration, 30 datapoints were accumulated at each time interval with both temporal profiles consisting of a minimum of 46 time intervals. This procedure was repeated for each atomic nitrogen concentration for a minimum of nine different atomic nitrogen concentrations. As CN and OH signals were potentially non-zero in the absence of the photolysis laser (due to the possible upstream production of CN and OH radicals by the microwave discharge) several time points were recorded by firing the probe lasers before the photolysis laser. The LIF intensities were measured at a fixed distance from the Laval nozzle; the chosen distance corresponded to the maximum displacement from the nozzle for optimal flow conditions to be still valid thus allowing us to exploit the fluorescence signals over as large a period as possible. Simple exponential fits to the OH and CN temporal profiles yielded the pseudo-first order rates for reactions 1 and 4, $k'_1$ and $k'_4$ respectively. Nevertheless, it should be noted that as $k'_1$ and $k'_4$ contain the additional loss terms $k_{L,OH}$ and $k_{L,CN}$, the losses of OH and CN respectively by diffusion from the zone illuminated by the probe laser (which is smaller than the cross sectional area of the supersonic flow) or through secondary reactions, we do not expect the fits to pass through the origin. In the absence of atomic nitrogen, this y-axis intercept value is equal to $k_{L,CN}$ - m $k_{L,OH}$ where m is the slope. As a result of the increasing value of m, the intercept value is expected to become more negative at higher temperatures.

**Chemical model.** To simulate the interstellar chemistry, the Nautilus gas-grain model was used which allows us to solve numerically a series of coupled differential equations. This model computes the species abundances as a function of time for a given set of physical parameters, starting from an initial chemical composition. A chemical network of 4394 gas-phase reactions and 1748 grain-surface and gas-grain reactions was used with values recommended by the KIDA database with the exceptions being reactions 1, 2 and 4. A more detailed description of the model itself can be found in Semenov et al. (38). This model has been used to study the chemistry in several types of environments such as dark molecular clouds (20) and protoplanetary disks (39, 40). In our model, ammonia is formed on grains by the successive hydrogenation of nitrogen atoms that stick on grain surfaces through the Langmuir-Hinshelwood mechanism (41).

For the prestellar core models, we have taken the example of Barnard 68, whose physical structure has been well studied (27, 42). The density structure is well reproduced by a Bonnor Ebert sphere with a central density of $5 \times 10^5$ cm$^{-3}$ (27). We have used Nautilus in a 1D geometry with this density structure and assumed a constant temperature of 10 K, following Maret et al. (6). The visual extinction is computed as a function of the radius to the centre of the core by integrating the $H_2$ column density. The penetration of the external UV photons (using the standard UV interstellar field) is computed as a function of this visual extinction. The chemical composition of the core is then computed as a function of time assuming that the physical structure does not change during this time. As for dense clouds, we have assumed that species are initially in atomic form, except for $H_2$.

**ACKNOWLEDGEMENTS.** The experimental work was supported by the Agence Nationale de la Recherche (ANR-JC08_311018), the Conseil Régional d'Aquitaine (20091102002), the European Union (PERG03-GA-2008-230805) and the CNRS interdisciplinary program EPOV. The experimental and modeling work were both supported by the INSU-CNRS national program PCMI and the Observatoire Aquitain des Sciences de l'Univers.

Author Contributions: K.M.H, V.W and A.B. designed research. J.D., K.M.H., A.B., M.C., J.-C.L., U.H. and V.W. performed research. K.M.H. and V.W. wrote the paper.
The authors declare no conflict of interest.
This article is a PNAS Direct Submission.
*To whom correspondence should be addressed. E-mail: km.hickson@ism.u-bordeaux1.fr.
This article contains supporting information online at www.pnas.org

**Figures**

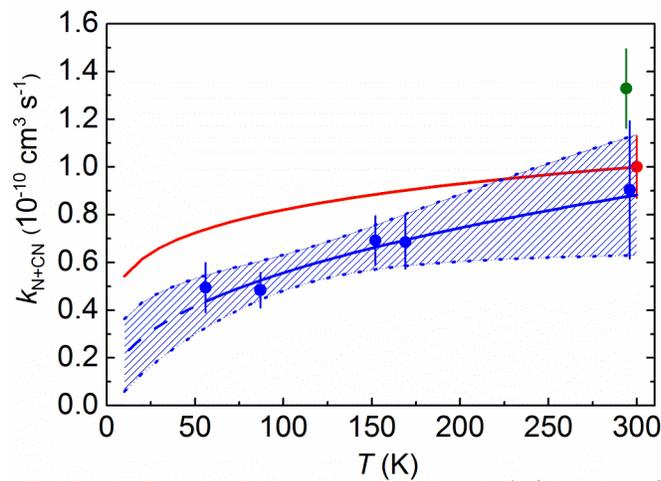

**Figure 1. Rate constants for the $N(^4S^0) + CN(^2\Sigma^+)$ reaction as a function of temperature.** Experimental values: (●) Whyte & Phillips (15); (●) Atakan et al. (16); (●) This work. Error bars on the current measurements represent the $1\sigma$ statistical uncertainty combined with a systematic uncertainty of 10 % estimated from possible inaccuracies in the measured flow rates, pressures and temperatures. ▬--- Fit and extrapolation to 10 K of the current data. The shaded area represents 95% confidence intervals for the rate constant fit given by a standard method of propagation of variances and covariances. Recommended values: (▬) KIDA (17).

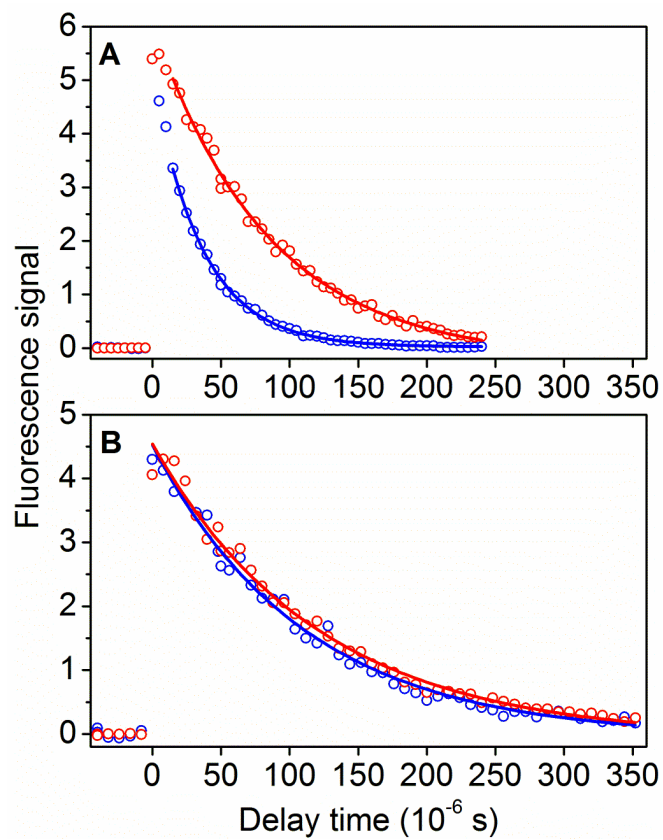

**Figure 2. Exemplary temporal profiles of OH and CN.** (**A**) At 296 K with estimated [N] = 2.4 × $10^{14}$ atom $cm^{-3}$. ○ CN($^2\Sigma^+$) LIF signal. ○ OH($^2\Pi_{3/2}$) LIF signal. (**B**) As in (**A**) but at 56 K with estimated [N] = 1.1 × $10^{14}$ atom $cm^{-3}$.

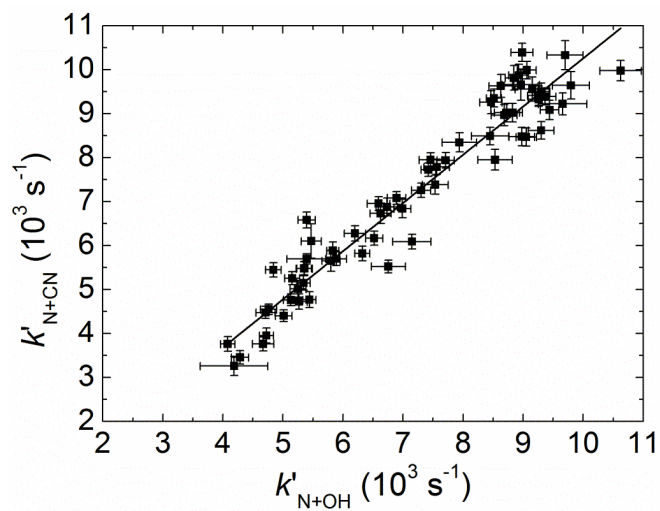

**Figure 3. Pseudo-first-order rate constants for reaction 4 as a function of the pseudo-first-order rate constants for reaction 1 at 56 K.** A weighted linear least squares fit yields the ratio of the second-order rate constants $k_4/k_1$. The error bars on the ordinate reflect the statistical

uncertainties at the level of a single standard deviation obtained by fitting to OH LIF profiles such as those shown in Fig. 2. The error bars on the abscissa were obtained in the same manner by fitting to the CN LIF profiles.

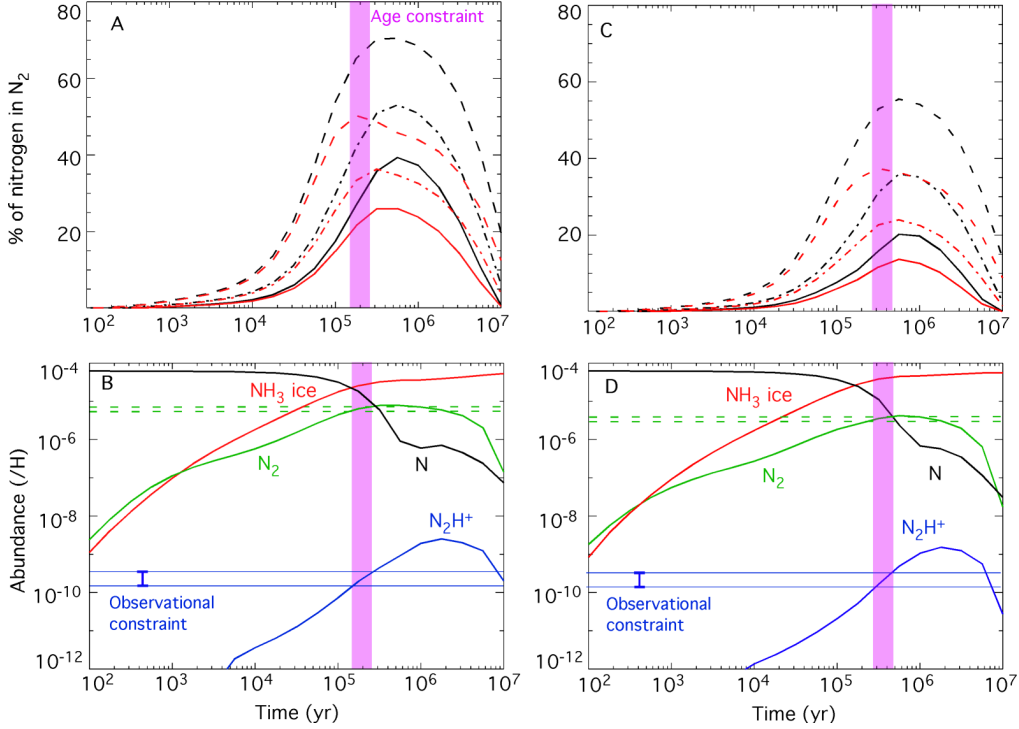

**Figure 4. Model results for the formation of selected nitrogen containing species in dark clouds as a function of time.** (**A**) Percentage of elemental nitrogen in the form of $N_2$ (in the gas-phase and on interstellar ices). Typical dense cloud parameters are used for a density of $2 \times 10^4$ cm$^{-3}$, a solar elemental abundance of $6.2 \times 10^{-5}$ (/$n_H$) for nitrogen and a C/O ratio of 1.2, for three different networks: dashed lines, model a) OSU database; dashed-dotted lines, model b) KIDA values (17); solid lines, model c) KIDA with updates for reactions 1, 2 and 4. Red lines indicate gas-phase $N_2$, black lines indicate total (gas + ice) $N_2$. (**B**) Gas-phase abundances of N, $N_2H^+$ and $N_2$ and abundance of $NH_3$ ice (/$n_H$) predicted by model c) for a C/O ratio of 1.2 together with the observational constraints on $N_2H^+$ in TMC-1 (5, 23). Horizontal dashed lines indicate the constraints on $N_2$ abundances. (**C**) As (**A**), for a C/O ratio of 0.7. (**D**) As (**B**), for a C/O ratio of 0.7.

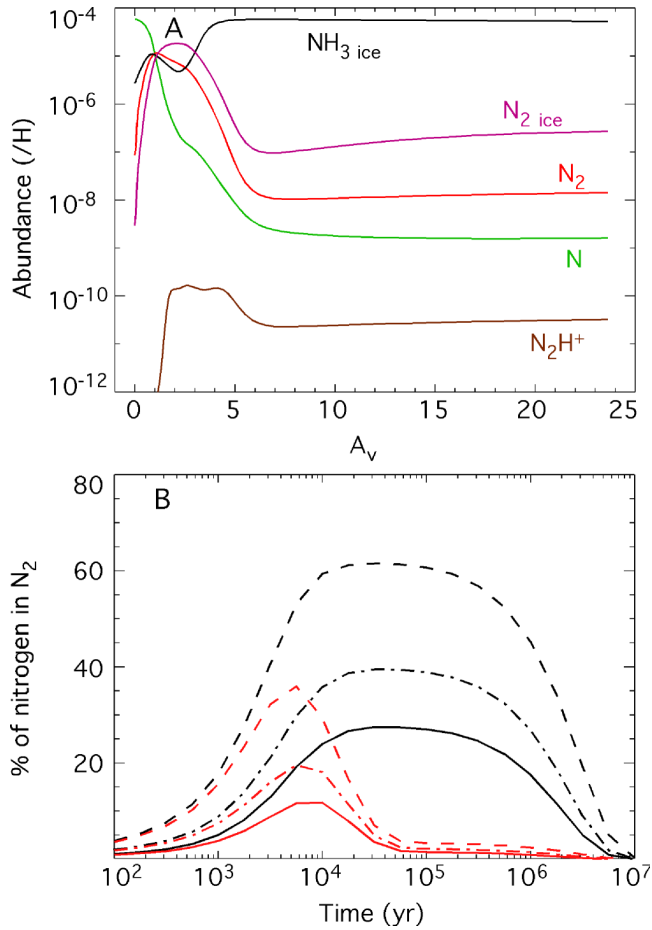

**Figure 5. Model results for the B68 prestellar core using a C/O elemental ratio of 1.2. (A)** Gas phase abundances of N, $N_2$, $N_2H^+$ and ice abundances of $N_2$ and $NH_3$ ($/n_H$) as a function of visual extinction for model c) at an age of $5 \times 10^6$ yr. **(B)** Percentage of elemental nitrogen in the form of $N_2$ (in the gas-phase and on interstellar ices) as a function of time. Red lines indicate gas-phase $N_2$, black lines indicate total (gas + ice) $N_2$ with three different networks: dashed lines, model a) OSU database; dashed-dotted lines, model b) KIDA values (17); solid lines, model c) KIDA with updates for reactions 1, 2 and 4.

**Supporting Information**

**SI Text**

**Experimental cross-checks.** Early test experiments at 152 K showed that the pseudo-first-order decay curves of the OH radicals such those shown in Fig. 2 did not vary as a function of the ICN concentration. In contrast, it was seen that under certain conditions, the CN radical fluorescence signal did not decay to zero at long times, but instead to a non-zero baseline value. This residual signal varied as a function of the gas-phase $H_2O_2$ (or $H_2O$) concentration, so that at high concentrations of $H_2O_2$, the CN fluorescence signal decayed to the pre-trigger zero level. At low $H_2O_2$ concentrations, the baseline value was seen to be as large as 9 % of the peak signal amplitude. When the values of $k'_4$ determined from fits to decays measured in the presence of low $[H_2O_2]$ were plotted as a function of the corresponding $k'_1$ value, the ratio $k_4 / k_1$ was seen to be larger than the ratio determined at high $[H_2O_2]$. Several test experiments under conditions identical to those used in these early experiments were performed to determine the origin of this effect. Using a solar blind channel photomultiplier tube (Perkin Elmer MP1911) in photon counting mode coupled to a monochromator with a grating blazed at 121 nm, it was possible to observe emission from excited $A^3\Sigma_u^+$ molecular nitrogen $N_2^*$ within the supersonic flow at several wavelengths between 145 and 185 nm. The emission intensity was seen to vary strongly as a function of $[H_2O_2]$ within the flow, with little or no emission occurring at the highest $[H_2O_2]$.

In light of these findings, further test experiments employing $CBr_4$ as a precursor for $C(^3P)$ atoms were performed at 152 K. Here, $CBr_4$ was entrained in the supersonic flow and underwent pulsed multi-photon dissociation at 266 nm. No other minor reagent precursor molecules were used, atomic nitrogen was produced in the usual manner (thereby producing $N_2^*$) and the probe laser was tuned to a CN $B^2\Sigma^+$ - $X^2\Sigma^+$ transition. Under these conditions, formation of CN radicals was clearly observed in the supersonic flow. Moreover, the rate of production of CN was seen to vary as a function of the atomic nitrogen concentration (by varying the $N_2$ flow through the discharge and/or the discharge power). This last observation indicates that the rate of formation of CN is strongly linked to $[N_2^*]$. We postulate that the following process may have led to secondary CN formation:

$$C(^3P) + N_2^* \rightarrow CN + N \qquad\qquad 5$$

with atomic carbon being formed as the primary product of reaction 4. Given the experimental evidence, two possible mechanisms are potentially responsible for inhibiting reaction 5, leading to zero residual baseline values. Firstly, it is probable that $N_2^*$ is either quenched or reacts with $H_2O_2$ or $H_2O$ given the dependence of the emission intensity on $[H_2O_2]$ or $[H_2O]$. Secondly, we cannot rule out the possibility that $C(^3P)$ is also removed by reaction with $H_2O_2$. In both cases, high $[H_2O_2]$ would lead to less secondary CN formation. As a result, all experiments used in the final analysis were those conducted with $[H_2O_2]$ greater than $3.9 \times 10^{12}$ cm$^{-3}$.

Although reaction 4 from ground state reactants $N(^4S^0)$ and $CN(^2\Sigma^+)$ leading to ground state products $C(^3P)$ and $N_2(^1\Sigma_g^+)$ is exothermic by 191.4 kJ mol$^{-1}$, the formation of excited state $C(^1D)$ is also energetically possible (although the products no longer correlate adiabatically with the reactants), with an exothermicity of 69.3 kJ mol$^{-1}$. As a result, further test experiments were performed at 152 K to check for the presence of these atoms. Unlike $C(^3P)$, $C(^1D)$ is known to react rapidly with $H_2$ at room temperature to form methylidyne radicals (S1), CH, which should themselves react with N($^4$S$^0$) to reform CN radicals (reaction 3). If present, the density of CH radicals should peak at a time delay governed by the ratio of the rate constants for the formation and loss processes. We looked for CH radicals over a range of time delays and over a range of atomic nitrogen concentrations by exciting the R1(1) line of the (1, 0) band of the B$^2\Sigma^-$ ← X$^2\Pi_{1/2}$ transition at 363.080 nm, and observing fluorescence around 390 nm via the (0, 0) band of the B$^2\Sigma^-$ → X$^2\Pi_{1/2}$ transition. Nevertheless, no fluorescence signal was ever seen at and around this wavelength indicating that C($^1$D) formation is negligible under our experimental conditions.

**Model cross-checks.** The typical elemental abundance ratio (C/O) used by many authors is less than one so that a large amount of atomic oxygen is still available for the chemistry once CO has been formed. Depletion of oxygen compared to typical atomic abundances used has been proposed by Jenkins (21) based on the observation of atomic lines towards hundreds of lines of sight with different densities. Whittet (22) studied the budget of O-bearing species as a function of cloud density. In Fig. 3 of his article, he represented the fraction of the cosmic oxygen abundance that was observed in refractory grain cores, in the gas-phase (mainly in the form of CO) and in the ices. If one removes all these components, a large fraction of the oxygen is in an unidentified form which does not seem to participate to the chemistry. Hincelin et al. (20) used this hypothesis to explain the low O$_2$ gas-phase abundance in dense clouds. Considering the uncertainty on the oxygen elemental abundance to be used in dense clouds, two different values have been used for oxygen, leading to C/O elemental ratios of 0.7 and 1.2. All the model runs that we performed showed that the use of a low C/O elemental abundance ratio led to lower N$_2$ abundances as explained by the influence of the O + CN reaction in the main article.
Other parameters can be important for our results. For example, the choice of the initial conditions (initial species abundances). In our typical model, we start with all the species in atomic form, except for H$_2$, because we assume that dense clouds form from diffuse media where the elements are mostly in the atomic form. Nevertheless, molecules such as CO, OH and CH, are observed in diffuse clouds with abundances around 10$^{-6}$ for CO and 10$^{-8}$ (/n$_H$) for OH and CH (S2, S3). The inclusion of these molecules in our models as initial conditions at their diffuse cloud abundances does not change the results for dense clouds.

The cosmic-ray ionization rate is another parameter which might change the model results. We tested its importance by running the model with two extreme values of the cosmic-ray ionization rate $\zeta$: 10$^{-18}$ and 10$^{-16}$ s$^{-1}$. The predicted abundances of the major N-bearing species as a function of time computed with these values are shown in Fig. S3, for dense cloud conditions and depleted oxygen (C/O = 1.2). The age for which the model agrees with N$_2$H$^+$ observations in TMC-1 (CP) is clearly very dependent on this parameter. For a larger $\zeta$ than typically used (10$^{-17}$ s$^{-1}$), the agreement between observations and models is obtained sooner (3-5 × 10$^4$ yr) and the predicted abundance of N$_2$ at that time is smaller. A smaller cosmic-ray ionization rate produces an agreement for larger times (8-10 × 10$^5$ yr) and again the predicted abundance of gas-phase N$_2$ is smaller than for our typical model results.

We show in Fig. S1 (**B**), the gas-phase abundances of CN and NO (/n$_H$) predicted by our model c) for the two C/O elemental ratios. The CN abundance observed in TMC-1 (CP) is reasonably reproduced by our models, at the times defined by the N$_2$H$^+$ observations, considering the observed values of 6 × 10$^{-8}$ (/n$_H$) of Ohishi (S4) and Crutcher et al. (S5), although a much smaller CN abundance of about 10$^{-9}$ (/n$_H$) has been observed by Pratap et al. (23). The NO abundance observed in the same cloud by Gerin et al. (S6) of $2.7 \times 10^{-8}$ ($/n_H$) is also in agreement with our models.

The results of the prestellar core model (B68) using the C/O elemental ratio of 1.2 are shown in Fig. 5 and the ones for the C/O ratio of 0.7 are presented in Fig. S2, both for a time of $5 \times 10^6$ yr. This time was selected to reproduce the observed abundance of $N_2H^+$ in B68 (6). At early times, the model produces too much $N_2H^+$ whereas at later times, it produces too little of it. This time is very different from the one determined by Maret et al. (6) of $2 \times 10^5$ yr, but in their model all of the carbon was initially present as CO and the remaining oxygen was in the form of water ice so that their timescale cannot be compared directly with ours. A few $10^6$ yr for the "chemical" age of a prestellar core is in agreement with the conclusion of Pagani et al. (S7) using deuterated species. In the case of prestellar cores, the choice of initial conditions can be debated since the material probably spent some time at intermediate densities during which some species are created. We tested this hypothesis by using the composition for a dense cloud computed by our model as the initial conditions for the prestellar core. We also ran simulations with half of the carbon already present as CO, keeping the rest in the atomic form. For these models, the results were similar to those shown in Figs. 5 and S3, however, the timescale to reach the observed $N_2H^+$ abundance was simply shifted towards smaller times. The main conclusion of this paper, that the use of up-to-date gas-phase chemistry decreases the fraction of nitrogen converted into $N_2$ at reasonable ages for both dense clouds and pre-stellar cores appears to hold even for a wide range of initial model parameters.

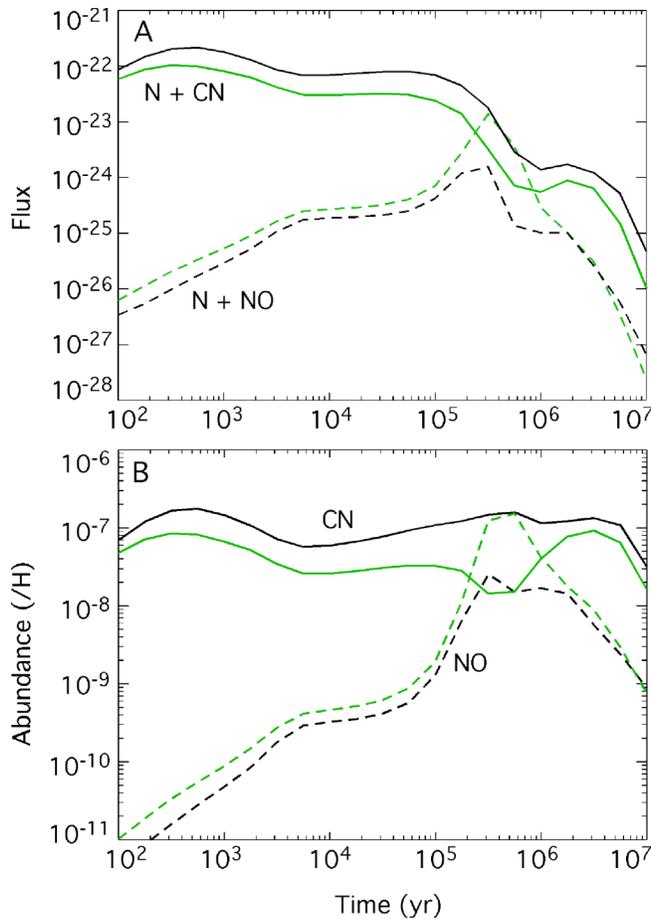

**Figure S1. Dense cloud model results for CN and NO molecules as a function of time.** (**A**) Fluxes (the product of the rate constants and the densities of the two reactants) for the reactions N + NO and N + CN for dense cloud conditions using model c) and for elemental abundance ratios C/O of 1.2 (black lines) and 0.7 (green lines). (**B**) Gas-phase abundances of CN and NO (/$n_H$) with the same models as (**A**).

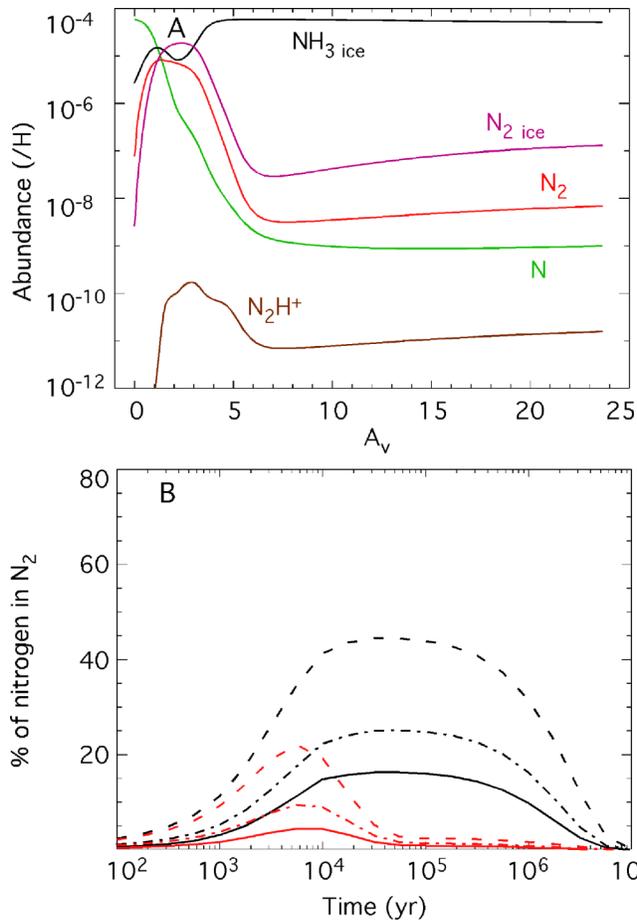

**Figure S2. Model results for the prestellar core B68 using a C/O elemental ratio of 0.7. (A)** Gas phase abundances of N, $N_2$, $N_2H^+$ and ice abundances of $N_2$ and $NH_3$ (/$n_H$) as a function of visual extinction for model c) at an age of $5 \times 10^6$ yr. **(B)** Percentage of elemental nitrogen in the form of $N_2$ (in the gas-phase and on interstellar ices) as a function of time. Red lines indicate gas-phase $N_2$, black lines indicate total (gas + ice) $N_2$ with three different networks: dashed lines, model a) OSU database; dashed-dotted lines, model b) KIDA values (17); solid lines, model c) KIDA with updates for reactions 1, 2 and 4.

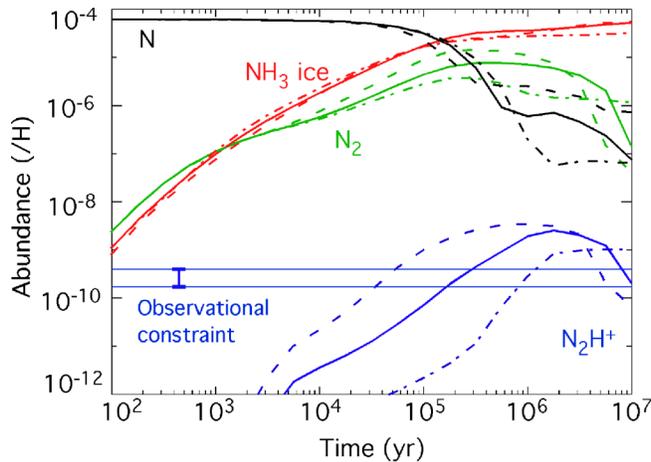

**Figure S3. Effect of the cosmic ray ionization rate on the fractional abundances of gas-phase N, $N_2$ and $N_2H^+$ and $NH_3$ ice (/$n_H$).** Dense cloud physical conditions are used (total density = $2 \times 10^4$ cm$^{-3}$) with different cosmic ray ionization rates: $10^{-18}$ s$^{-1}$ (dashed-dotted lines), $10^{-17}$ s$^{-1}$ (solid lines) and $10^{-16}$ s$^{-1}$ (dashed lines). Horizontal blue lines represent the observational constraints on $N_2H^+$ from TMC-1 (CP) (5, 23).

**Supplementary References**